# Is it possible for a perovskite p-n homojunction to persist in the presence of mobile ionic charge?

Philip Calado and Piers R. F. Barnes
Department of Physics, Imperial College London, London, SW7 2AZ
Email: p.calado13@imperial.ac.uk, piers.barnes@imperial.ac.uk

Recently Cui et al.[1] reported on the fabrication a p-n homojunction perovskite solar cell (PSC) using stoichiometric control of sequentially-deposited perovskite layers. The authors propose that the junction leads to an enhanced electric field in the perovskite absorber resulting in improved charge separation. In this work we show that the data presented in the paper does not directly support this claim. Furthermore, Cui et al.'s thesis is not compatible with the large body of existing literature showing that mobile ionic defects present in methyl-ammonium lead iodide (MAPI) and its derivatives are highly mobile at room temperature. Using drift diffusion device simulations we show that large densities of mobile ionic charge in the system are likely to the screen any beneficial effects of a p-n homojunction.

A comparison of the electric field strength difference calculated from the cross-sectional Kelvin probe force microscopy (KPFM) data presented by Cui et al. for the homojunction versus the undoped control device (reproduced in Supplementary Figure 1) indicates a field reduction, by approximately a factor of 2 at forward bias, rather than an enhancement as suggested on pp. 154. While a direct comparison of these data is already questionable since the undoped control sample included transport layers not present in the homojunction KPFM sample (despite the homojunction solar cell using transport layers), the data does not support their claim that the electric field is enhanced by the homojunction. Additional issues regarding the conclusions drawn from the experimental data are discussed in Supplementary Note 1.

We now consider how mobile ionic charge would affect the electric field at a junction formed between p- and n-type perovskite materials. There is overwhelming experimental and theoretical evidence that high concentrations of mobile ionic defects are present in lead-halide perovskites.[2–5] Further references to support this position can be found in Supplementary Note 2. This includes PSCs that do not show current-voltage (*J-V*) hysteresis at standard laboratory scan rates (~1 $mVs^{-1}$ to ~100 $mVs^{-1}$).[6–8] In Cui et al. Figure 5b the photocurrent of a homojunction is shown at an applied potential of $V_{app}$ = 0.95 V. The ~20 s stabilisation time for the current and the *J-V* hysteresis in Cui et al. Figure 5d are classic signatures of slow moving mobile ionic charge in the perovskite absorber.[4,9,10] While the workfunction of isolated perovskite films may be modified by choice of the precursor stoichiometry (Cui et al. Figure 3b), the initial differences are unlikely to be maintained once the materials are brought into contact with one another and interdiffusion of ionic defects takes place. Although the Transmission Electron Microscope (TEM) data presented in Cui et al. Supplementary Figure 20 indicates that bromine has not diffused significantly into the $MAPbI_3$ phase, this does not preclude the transport of iodide vacancies across the interface. Furthermore, accumulation of ionic charge at the surface of the different perovskite materials would alter the electrostatic potential at the surface, causing a shift in the measured workfunction as measured by X-ray Photoelectron Spectroscopy (XPS), which could be incorrectly interpreted as a shift in the electronic band occupancy (chemical potential shift). Certainly these two effects cannot be easily decoupled.

Previous works suggest that PSCs can only be accurately simulated by models that include mobile ionic charge.[7,11–15] As such, it is unlikely that the simulations presented by Cui et al. (Cui et al. Figure 2c and Supplementary Figure 4) meaningfully predict the physics of their devices. To test the possibility that charge separation could be enhanced by the built-in field of a perovskite homojunction we simulated devices with and without mobile ions using our open source device simulation tool Driftfusion.[16,17] We consider four doping schemes in the perovskite absorber layer: 1. An intrinsic (undoped) layer throughout; 2. A p-n homojunction; 3. n-type throughout (donor density, $N_D$ = 3 x $10^{17}$ $cm^{-3}$); 4. A p-n junction with *mobile* ionised donors ($N_D$ = 3 x $10^{17}$ $cm^{-3}$). In all devices the perovskite layer is sandwiched by electron and hole selective contacts simulating $TiO_2$ and spiro-OMeTAD respectively. Where additional mobile ions are simulated, charge-neutral Schottky defects with a density of $N_{ion} = 10^{19}$ cm$^{-3}$ [18] were included in addition to any static dopant charges in the perovskite layer. A lower density of $N_{ion} = 10^{17}$ cm$^{-3}$ was also tested (Supplementary Figures 4 and 5) since factors such as lattice strain may influence the defect density.[19] Here, positively charged mobile ionic charges with a mobility of $10^{-10}$ cm$^2$ V$^{-1}$ s$^{-1}$ are confined to the perovskite layer with a uniform background of static negative counter ions. In the 'mobile donors' case donor dopants were given a mobility of $10^{-10}$ cm$^2$ V$^{-1}$ s$^{-1}$ with the Schottky defect density set to zero. Further details of the simulations are given in the Supplementary Information Simulation Methods.

Figure 1a shows maximum power point current densities for the intrinsic and p-n homojunction devices with (solid curves) and without (dashed curves) mobile ionic charge. Only the simulations including mobile ions reproduce a slow transient current response similar to that seen in the experimental data (Cui et al. Figure 5b) suggesting that mobile ionic charge in the perovskite is necessary to accurately model the real system. The *J*-*V* scan results (full scan protocol given in the Supplementary Information Simulation Methods) for devices without mobile ionic charge ($N_{ion} = 0$ cm$^{-3}$) are given in Figure 1b and reproduce a $V_{OC}$ enhancement for the homojunction (blue curve) relative to the intrinsic absorber device (black curve) consistent with the simulations by Cui et al. (Cui et al., Supplementary Information Figure 5). The further enhancement in power conversion efficiency (PCE) with the n-type absorber is discussed below.

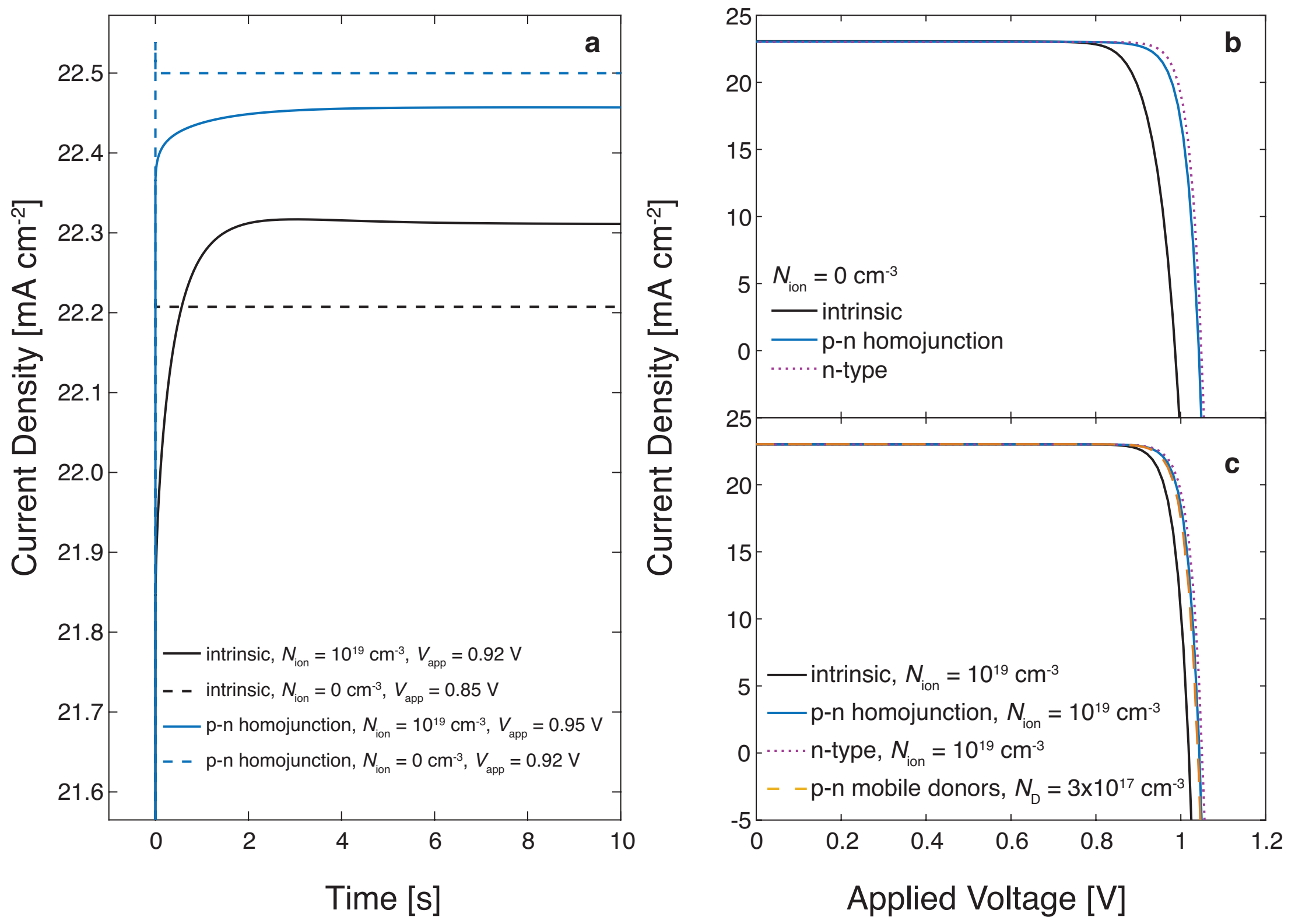

**Figure 1. Stabilised maximum power point current and current-voltage scans and for simulated perovskite devices. a**, Maximum power point current density as a function of time after switching from equilibrium to maximum power point under 1 Sun illumination. The applied voltage, $V_{app}$ in each case is given in the legend. Reverse scan, current-voltage characteristics for devices **b**, with zero or static dopants and **c**, mobile ionic charge in the perovskite absorber layer. Note that the 'mobile donors' simulations share the same *J*-*V* curve as that of the homojunction in **b**.

When mobile ions are included in the simulations (Figure 1c) the $V_{OC}$ for the intrinsic device improves by approximately 20 mV while the open circuit voltages for the other cases remain approximately unchanged ($V_{OC}$ = 1.04 – 1.05 V). To assess whether the p-n junction and an associated field is responsible for the improved $V_{OC}$ in the homojunction relative to the intrinsic device we analysed the mobile ionic charge density, electric field strength, electron and hole density, and recombination profiles of the four different doping schemes at approximately open circuit. In each instance mobile ions have redistributed in the active layer to screen out the electric field (Figure 2a). Only a small field persists in the homojunction and mobile donor cases (orange and blue curves respectively). The corresponding profiles without mobile ionic charge can be seen in Supplementary Figure 3.

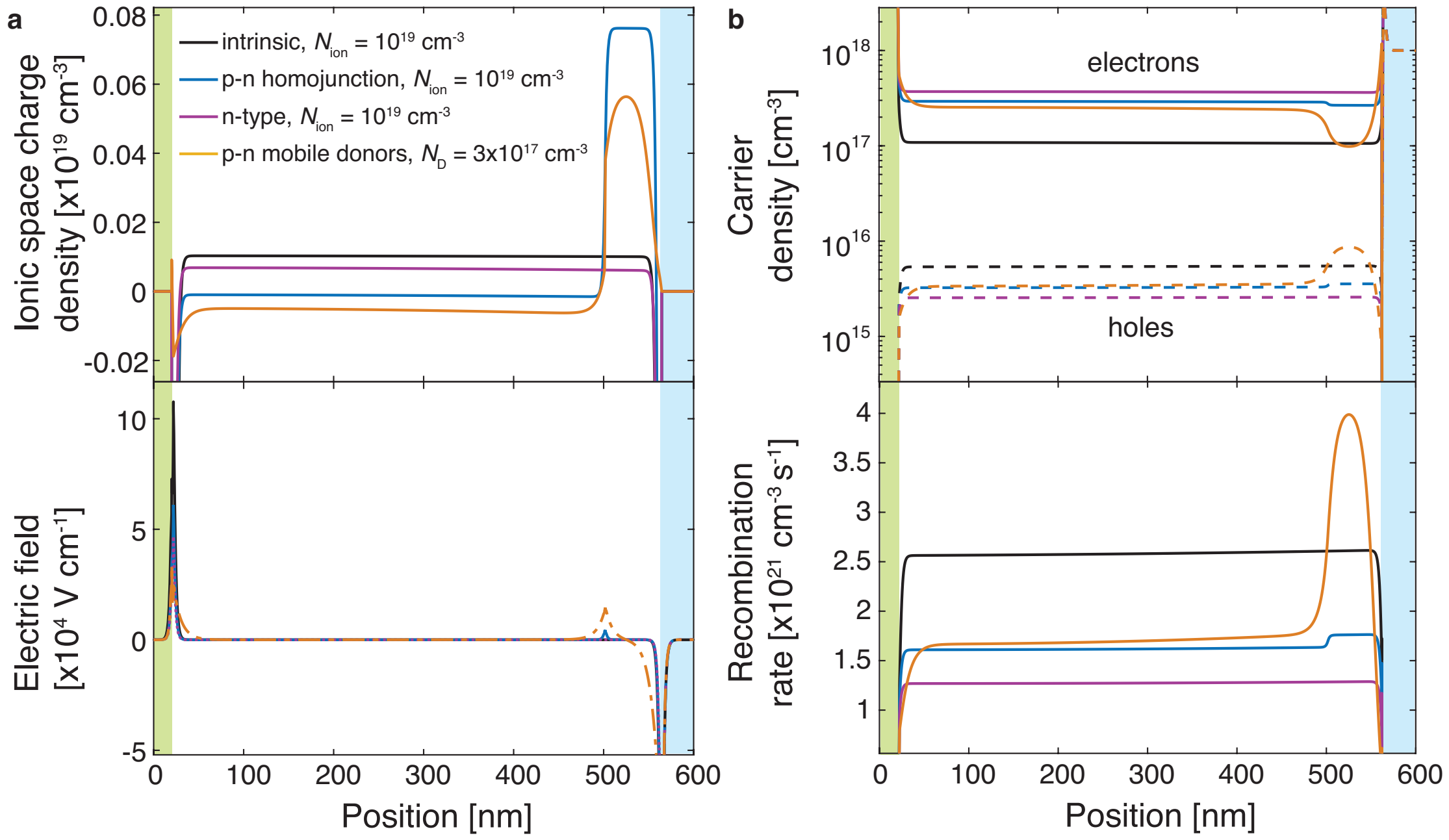

**Figure 2. Effect of mobile ions on charge carrier density, electric field strength, and recombination profiles. a,** Simulated ionic space charge and electric field strength profiles close to open circuit with 1 sun illumination. **b,** Corresponding free electron and hole density, and recombination rate profiles for the four absorber layer doping schemes (c.f. Figure 1b). $V_{app}$ = 1.02 V for the intrinsic device and 1.04 V for the three doped devices. In **b** solid curves indicate electron densities while dashed curves indicate hole densities. Titanium dioxide, perovskite and spiro-OMeTAD layers are indicated by green, white and blue regions respectively.

The carrier densities in Figure 2b indicate that any p-n doping is effectively screened by the redistribution of mobile ionic charge. The recombination-limiting factor in all cases is the hole density, evidenced by the observation that the SRH recombination rate profiles mirror those of the holes. Any enhancement to the $V_{OC}$ is therefore a direct result of minimisation of the minority carrier concentration in the perovskite layer. This accounts for why n-type doping of the perovskite in the simulated devices without ions leads to a greater enhancement of the $V_{OC}$ than the homojunction. We note that, in reality, doping may have a functional relationship to the density of non-radiative recombination centres and so could, instead be detrimental to device performance. Where the p-doping effect remains however, an increase in the minority carrier concentration (holes) and associated local increase in trap-mediated recombination at the junction ($x$ = 501 nm) at open circuit is observed. The same effect is seen both in devices without Schottky defects (Supplementary Figure 3) and with a lower Schottky defect density of $N_{ion}$ = $10^{17}$ $cm^{-3}$ (Supplementary Figures 4 and 5). This is counter to the principle proposition of Cui et al. that the p-n junction could be responsible for reduced trap-assisted recombination and an enhanced $V_{OC}$ (Cui et al., pp.152, paragraph 1).

To conclude, we address the probable origin of the improved PCE seen by Cui et al. in the homojunction device as compared to the standard devices (Cui et al. Figure 5c). It is well established that stoichiometry of the perovskite precursors plays an important role in the performance of the material.[20–23] The improved performance may arise from passivation of grain boundaries and recombination centres at transport layer interfaces and an associated reduction in non-radiative recombination[22,23] or owing to a doping effect throughout the entire perovskite layer as seen in our simulations. Given that approximately 90 % (480 nm of a 540 nm layer) of the p-n homojunction perovskite layer is fabricated using a $PbI_2$/MAI ratio of 1.10, the likely performance enhancement owing to the non-stoichiometry cannot be separated from any beneficial effects due to the proposed homojunction.

**Author Contributions**

PC initiated and performed the simulations, and drafted the manuscript. PC and PRFB contributed equally to the analysis.

**Acknowledgements**

We thank the UK Engineering and Physical Sciences Research Council (EPSRC) for funding this work (Grants: EP/J002305/1, EP/M025020/1, EP/M014797/1, EP/N020863/1, EP/R020574/1, EP/R023581/1, and EP/L016702/1).

**Competing interests statement**

The authors declare no competing interests.

**Is it possible for a perovskite p-n homojunction to persist in the presence of mobile ionic charge?**

Philip Calado and Piers R. F. Barnes
Department of Physics, Imperial College London, London, SW7 2AZ
Email: p.calado13@imperial.ac.uk, piers.barnes@imperial.ac.uk

## Supplementary Information

## Simulation Methods

Driftfusion[1] solves the coupled continuity equations for electrons, holes and a single ionic species and Poisson's equation in one dimension. A complete description of the device simulation methods can be found in Ref [2].

As far as possible we have used the parameters given in Cui et al Supplementary Tables 3 and 4.[3] Recombination parameters and models were not given in the paper or provided on request so our approach is as follows:

1. From the non-ideal current-voltage ($J$-$V$) curves given in Cui et al. Supplementary Figure 5 (ideality factor > 1) we assumed that first order bulk recombination is the dominant mechanism. See Ref [4] for rationale.
2. The radiative recombination coefficients were set according to the Shockley-Queisser limit for each semiconductor layer.[5,6]
3. Bulk Shockley Read Hall (SRH) recombination time constants were adjusted uniformly throughout the device to give approximately the same open circuit voltage ($V_{OC}$) as given in the simulations in Cui et al for devices without mobile ions.

Further assumptions owing to omitted parameters from Cui et al:

1. We have set the spiro-OMeTAD conduction and valence band effective density of states (eDOS) to $10^{20}$ $cm^{-3}$ rather than $10^{18}$ $cm^{-3}$. The simulations Cui et al. do not show a degenerate semiconductor (Supplementary Figure 4 shows a decay in carrier concentration inside the spiro-OMeTAD) as suggested by the device parameters in Cui et al. Supplementary Table 4 and Supplementary Table 5, where the doping and eDOS are equal.
2. We assume that the Fermi level of the metal electrodes, which define the electric potential boundary conditions of the system, are at the same level as those of the associated selective contact materials at equilibrium (i.e. the metal electrodes form Ohmic contacts with n-type $TiO_2$ and p-type spiro-OMeTAD).

The full set of parameters used in the device simulations can be found in Supplementary Table 1 and Supplementary Table 2.

We note that our simulations do not show the same accumulation of holes in the absorber layer and depletion of electrons in the Titanium dioxide ($TiO_2$) under short circuit conditions (Supplementary Figures 2a and 2b) as seen in Cui et al. Supplementary Figures 4c and 4d. The p-type nature of the perovskite layer in Cui et al may be due to the trapping model used for the simulations although since the details of the model have been omitted from the text we are unable to speculate further on its origin here.

Where devices are illuminated, the AM 1.5G solar spectrum is used to calculate a Beer-Lambert generation profile[2] using complex refractive index data from Ref [7].

## Simulation scan protocol

Devices are preconditioned at equilibrium. A cyclic voltage scan is then performed from 0 V – 1.2 V – 0 V at a scan rate of 1 mVs-1, sufficiently slow that the ionic distribution is close to a dynamic equilibrium.

## Kelvin probe force microscopy data

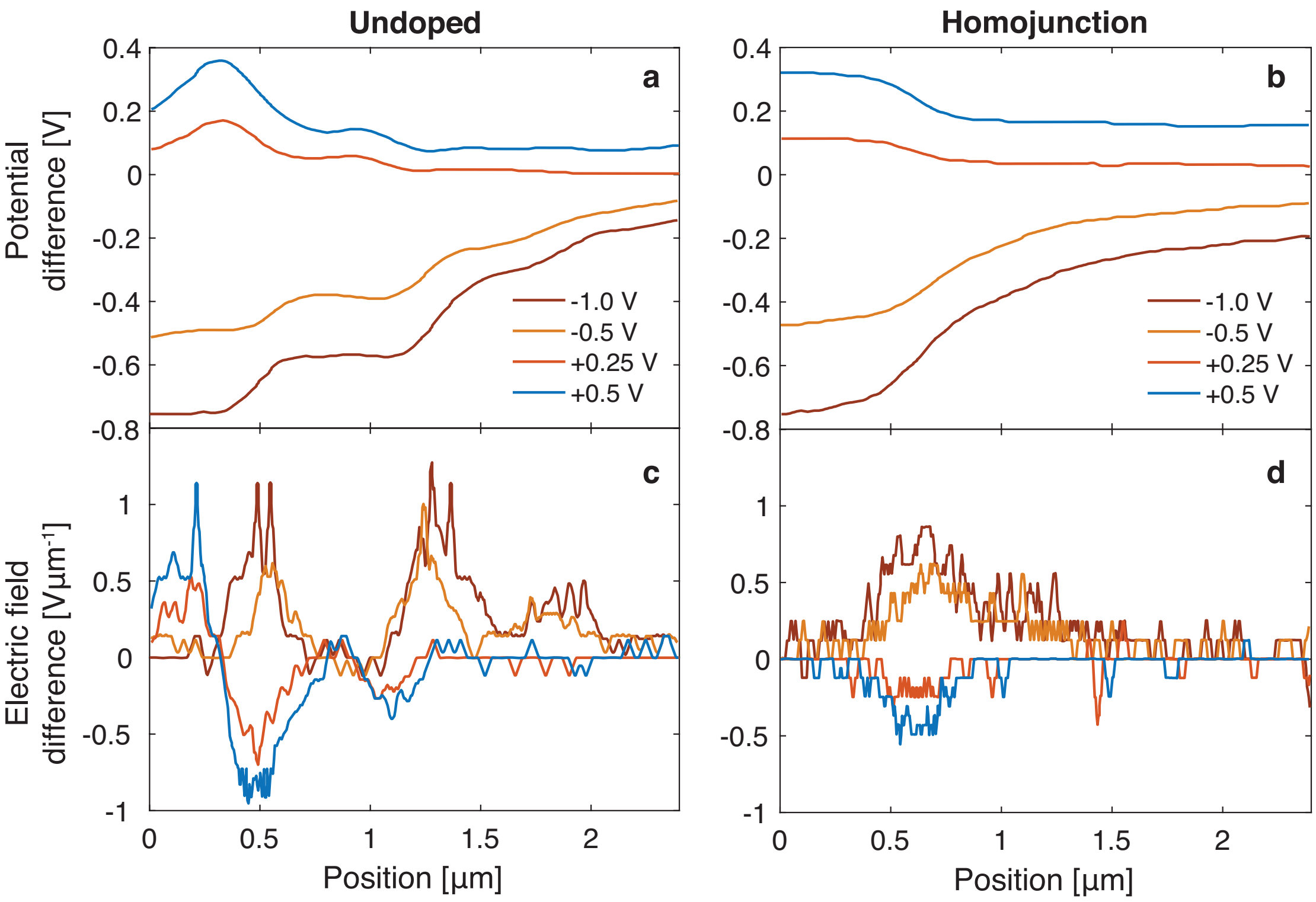

**Supplementary Figure 1. Kelvin probe force microscopy data showing the difference in electric potential from 0 V between the homojunction and the standard device.** Data in **a** and **b** reproduced from Cui et al. 2019.

## Simulated devices at short circuit ($V_b$ = 0 V), 1 Sun illumination

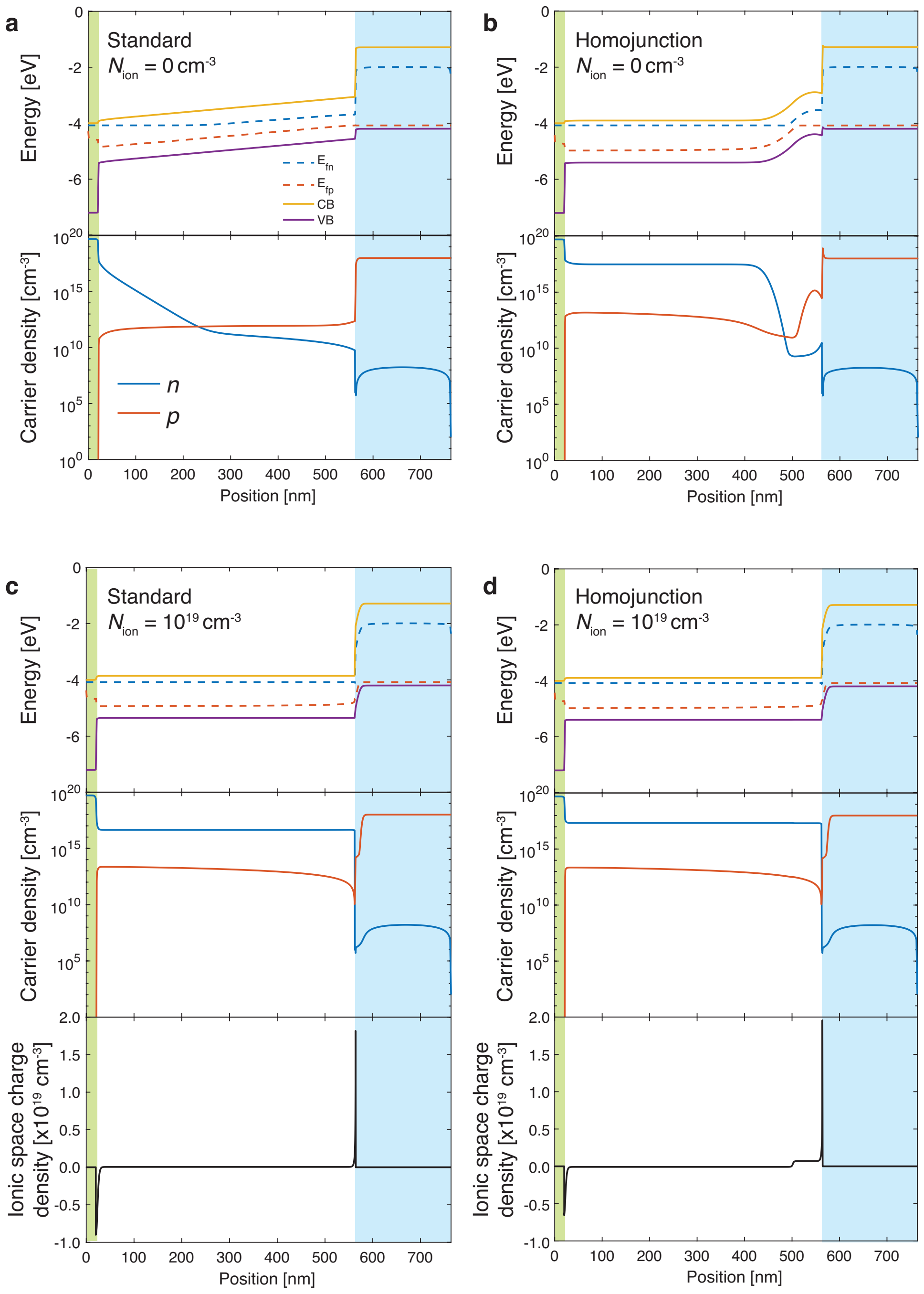

**Supplementary Figure 2. Energy level diagrams, carrier densities and mobile ionic charge densities for devices at short circuit under 1 sun illumination. a**, Standard p-i-n device without mobile ions ($N_{ion}$ = 0 cm$^{-3}$). **b**, Homojunction without mobile ions ($N_{ion}$ = 0 cm$^{-3}$). **c**, Standard p-i-n device with $N_{ion}$ = $10^{19}$ cm$^{-3}$. **d**, Homojunction with $N_{ion}$ = $10^{19}$ cm$^{-3}$. Titanium dioxide, perovskite and Spiro-OMeTAD layers are indicated by green, white and blue regions respectively.

## Device simulations at open circuit with $N_{ion}$ = 0 cm$^{-3}$

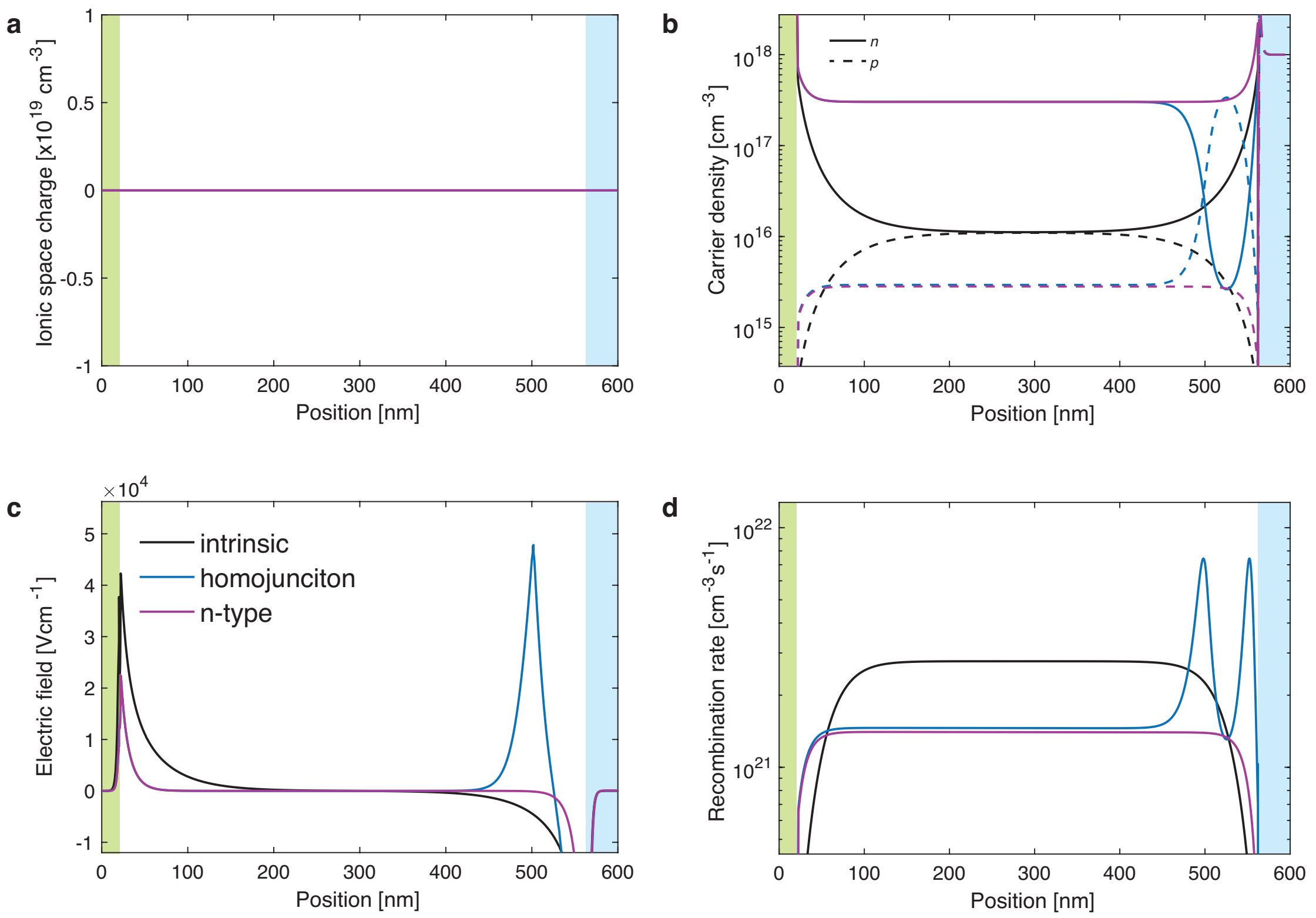

**Supplementary Figure 3. Ionic space charge distribution, electric field strength, electron and hole carrier density, and recombination rate for simulated devices without mobile ionic charge for four absorber layer doping schemes at approximately open circuit.**

## Device simulations with $N_{ion}$ = $10^{17}$ cm$^{-3}$

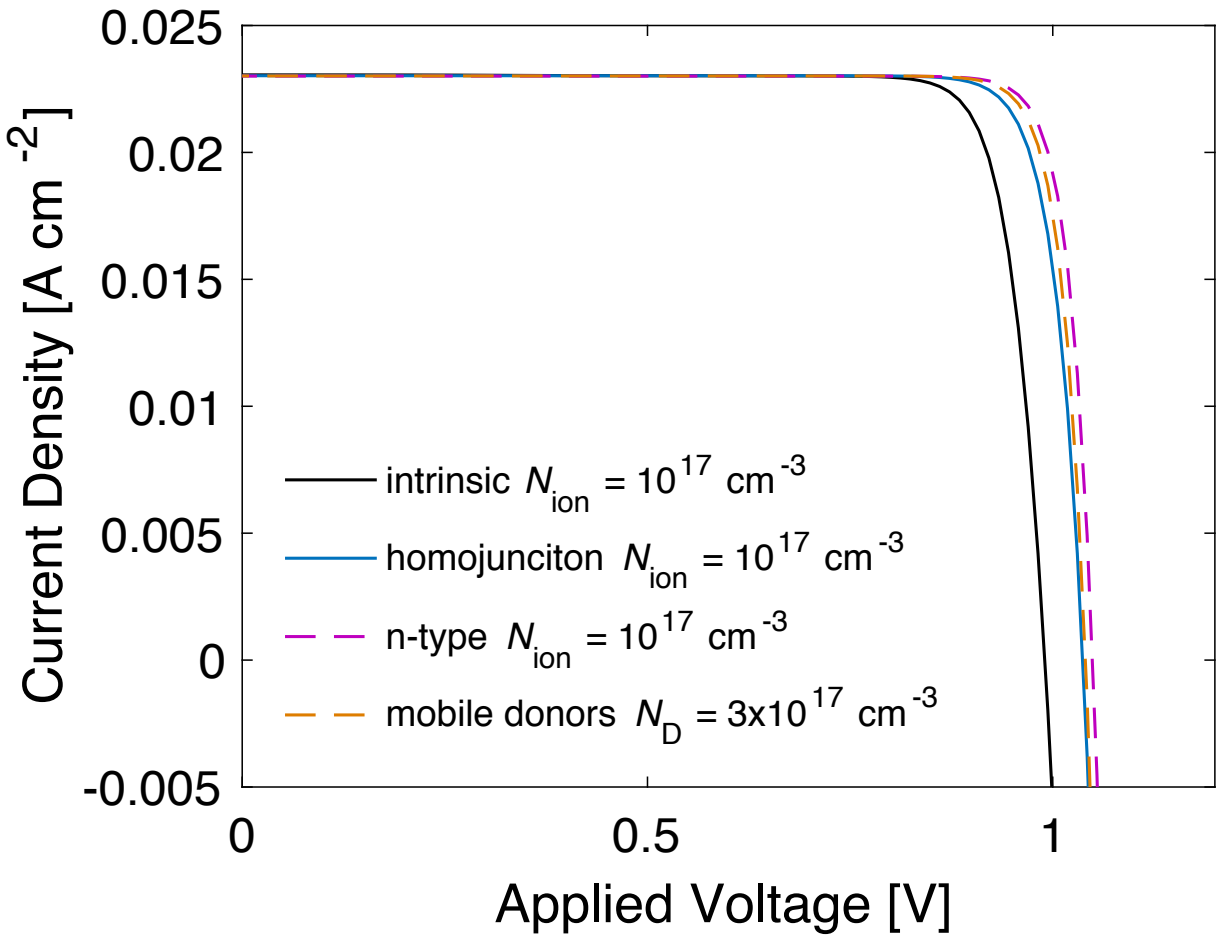

**Supplementary Figure 4. Current-voltage scans for simulated devices with $10^{17}$ cm$^{-3}$ mobile ions.**

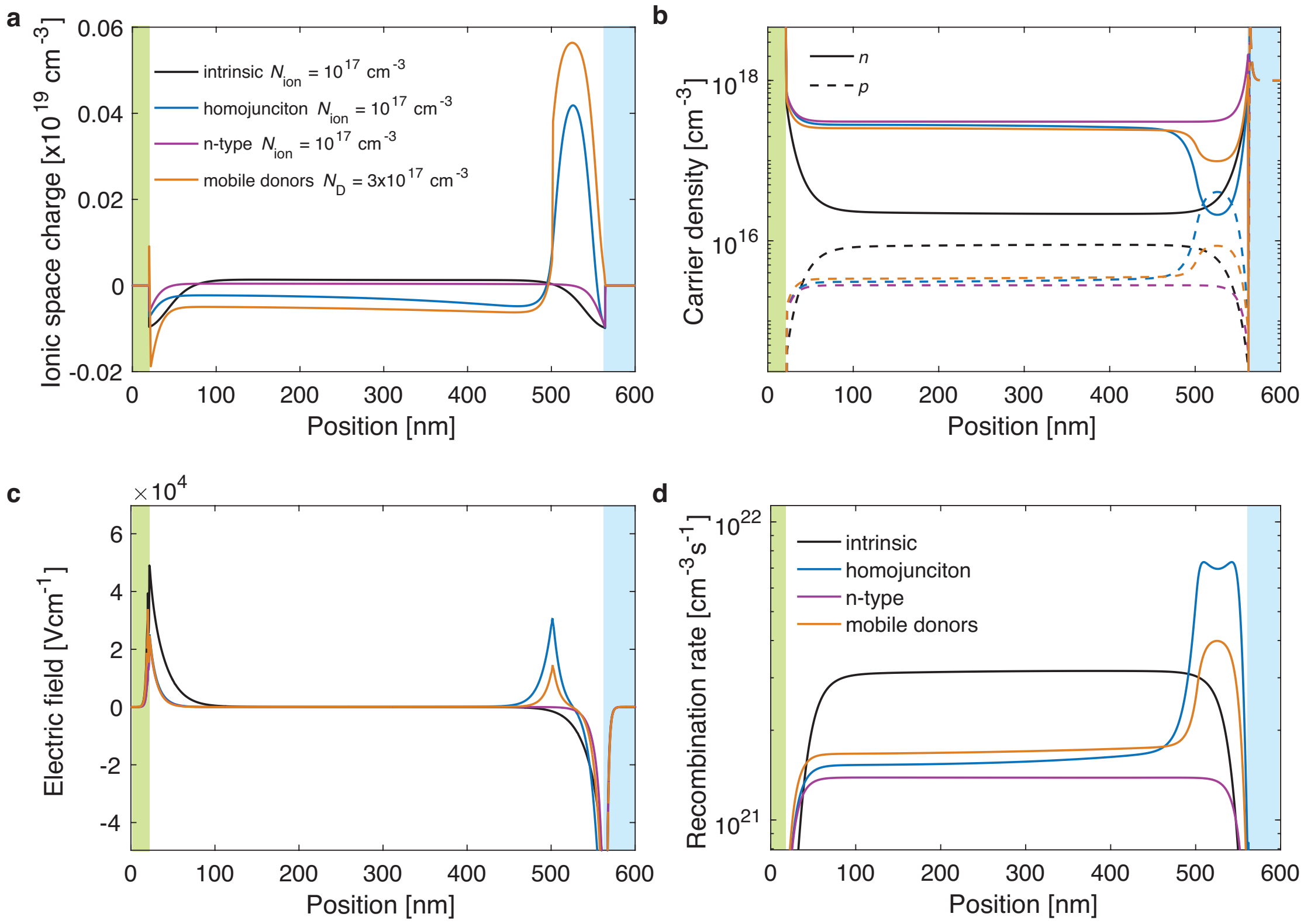

**Supplementary Figure 5. Ionic space charge distribution, electric field strength, electron and hole carrier density, and recombination rate for simulated devices including mobile ionic charge for four absorber layer doping schemes at approximately open circuit with $N_{ion}$ = $10^{17}$ cm$^{-3}$.**

## Supplementary note 1: Additional concerns with the experimental data

### XPS Workfunction measurements

The normalised XPS data for $PbI_2$/MAI = 1.05 and $PbI_2$/MAI = 1.1 appear to be very similar (Cui et al. Figure 3a, red and black curves respectively). The extrapolated workfunction shift here is highly dependent on the choice of the fitting range.

The schematics in Cui et al. Figure 3b show the conduction band at -3.9 eV yet the difference from the Fermi level for the three $PbI_2$/MAI ratios places it at -3.75 eV.

### KPFM Resolution

The spatial resolution of the KPFM measurement appears to limit the accuracy of estimates regarding location of the maximum field strength. This is evidenced by data showing an electric field penetrating >200 nm inside the Au regions of the devices under different bias conditions (Cui et al. Figure 4e and Supplementary Figure 12b). Since electric fields are not expected to extend beyond the surface of a metal, we assume that the observed field is an artefact arising from a combination of the tip size (25 nm), lateral averaging, and post-processing/smoothing of the data. Notwithstanding, the 2-dimensional KPFM data given in Supplementary Figures 10b – 10f does show a clear demarcation between p- and n-type perovskite layers. As the authors highlight, KPFM is a surface measurement and will be dominated by surface polarisation up to a depth of the screening length in any particular material. To mitigate this, the authors use a difference method and assume that the surface polarisation remains constant under different bias conditions.[8] However, we know that the devices take > 20 s to stabilise at maximum power point. This is a signature for large changes in the accumulation of mobile ionic charge in the device that are highly likely to simultaneously change the surface polarisation state of the cross-section.

### Time-resolved photoluminescence recombination model

Cui et al. apply a recombination model to time-resolved photoluminescence (TRPL) measurements on FTO/perovskite/PMMA trilayers to extract a trap assisted recombination lifetime in the homojunction versus an 'n-type' perovskite layer (Cui et al. Figure 3d). TRPL on perovskite/selective contact bilayers has previously been successfully used to measure the diffusion coefficient of charge carriers collected from perovskite layers[9][10][11] (i.e. charge transport). The application of a recombination model in this instance is therefore not appropriate.

## Supplementary note 2: Further references on ion migration in perovskites

Further evidence for ion migration in lead-halide perovskites can be found in references [12–18].

## Simulation parameters

| Property | Symbol | $TiO_2$ | Perovskite | | | | | | Spiro | Units |
|---|---|---|---|---|---|---|---|---|---|---|
| | | | Undoped | Homojunction | | n-type | Mobile donors | | | |
| | | | | n-type | p-type | | n-type | p-type | | |
| Layer thickness | $d$ | 20 | 540 | 480 | 60 | 540 | 480 | 60 | 200 | nm |
| Electron affinity | $\Phi_{EA}$ | -4.00 | -3.93 | -3.93 | -3.93 | -3.93 | -3.93 | -3.93 | -2.20 | eV |
| Ionisation potential | $\Phi_{IP}$ | -7.20 | -5.43 | -5.43 | -5.43 | -5.43 | -5.43 | -5.43 | -5.11 | eV |
| Equilibrium Fermi level | $E_0$ | -4.08 | -4.68 | -4.10 | -5.27 | -4.68 | -4.10 | -5.27 | -5.00 | eV |
| SRH trap energy | $E_t$ | -5.60 | -4.68 | -4.68 | -4.68 | -4.68 | -4.68 | -4.68 | -3.66 | eV |
| Conduction band effective density of states | $N_c$ | $10^{21}$ | $2.50 \times 10^{20}$ | $2.50 \times 10^{20}$ | $2.50 \times 10^{20}$ | $2.50 \times 10^{20}$ | $2.50 \times 10^{20}$ | $2.50 \times 10^{20}$ | $10^{20}$ | $cm^{-3}$ |
| Valence band effective density of states | $N_v$ | $2.00 \times 10^{20}$ | $2.50 \times 10^{20}$ | $2.50 \times 10^{20}$ | $2.50 \times 10^{20}$ | $2.50 \times 10^{20}$ | $2.50 \times 10^{20}$ | $2.50 \times 10^{20}$ | $10^{20}$ | $cm^{-3}$ |
| Donor dopant density | $N_D$ | $5.01 \times 10^{19}$ | - | $3.00 \times 10^{17}$ | - | $3.00 \times 10^{17}$ | $3.00 \times 10^{17}$ | - | - | $cm^{-3}$ |
| Acceptor dopant density | $N_A$ | - | - | - | $4.99 \times 10^{17}$ | - | - | $4.99 \times 10^{17}$ | $9.98 \times 10^{17}$ | $cm^{-3}$ |
| Shottky defect density | $N_{ion}$ | - | $10^{19}$ | $10^{19}$ | $10^{19}$ | $10^{19}$ | - | - | 0 | $cm^{-3}$ |
| Ionic site density | $N_{site}$ | - | $1.21 \times 10^{22}$ | $1.21 \times 10^{22}$ | $1.21 \times 10^{22}$ | $1.21 \times 10^{22}$ | - | - | 0 | $cm^{-3}$ |
| Electron mobility | $\mu_e$ | $6 \times 10^{-3}$ | 10 | 10 | 10 | 10 | 10 | 10 | $10^{-3}$ | $cm^2 V^{-1} s^{-1}$ |
| Hole mobility | $\mu_h$ | $6 \times 10^{-3}$ | 10 | 10 | 10 | 10 | 10 | 10 | $10^{-3}$ | $cm^2 V^{-1} s^{-1}$ |
| Ion mobility | $\mu_{ion}$ | 0 | $10^{-10}$ | $10^{-10}$ | $10^{-10}$ | $10^{-10}$ | $10^{-10}$ | $10^{-10}$ | 0 | $cm^2 V^{-1} s^{-1}$ |
| Relative dielectric constant | $\varepsilon_r$ | 100 | 25 | 25 | 25 | 25 | 25 | 25 | 3 | |
| Radiative recombination coefficient | $k_{rad}$ | $3.16 \times 10^{-14}$ | $4.79 \times 10^{-15}$ | $4.79 \times 10^{-15}$ | $4.79 \times 10^{-15}$ | $4.79 \times 10^{-15}$ | $4.79 \times 10^{-15}$ | $4.79 \times 10^{-15}$ | $2.99 \times 10^{-13}$ | $cm^3 s^{-1}$ |
| SRH electron time constant | $\tau_{n,SRH}$ | $2 \times 10^{-6}$ | $2 \times 10^{-6}$ | $2 \times 10^{-6}$ | $2 \times 10^{-6}$ | $2 \times 10^{-6}$ | $2 \times 10^{-6}$ | $2 \times 10^{-6}$ | $2 \times 10^{-6}$ | s |
| SRH hole time constant | $\tau_{p,SRH}$ | $2 \times 10^{-6}$ | $2 \times 10^{-6}$ | $2 \times 10^{-6}$ | $2 \times 10^{-6}$ | $2 \times 10^{-6}$ | $2 \times 10^{-6}$ | $2 \times 10^{-6}$ | $2 \times 10^{-6}$ | s |

## Supplementary Table 1. Layer properties used for the simulations.

| Property | Symbol | | Units |
|---|---|---|---|
| Electron surface recombination velocity (left boundary) | $s_{n,l}$ | $10^7$ | $cm\ s^{-1}$ |
| Hole surface recombination velocity (left boundary) | $s_{n,r}$ | $10^7$ | $cm\ s^{-1}$ |
| Electron surface recombination velocity (right boundary) | $s_{p,l}$ | $10^7$ | $cm\ s^{-1}$ |
| Hole surface recombination velocity (right boundary) | $s_{p,r}$ | $10^7$ | $cm\ s^{-1}$ |
| Series resistance | $R_s$ | 0 | Ohms |
| Workfunction (left boundary) | $\Phi_l$ | 4.08 | eV |
| Workfunction (right boundary) | $\Phi_r$ | 4.99 | eV |

## Supplementary Table 2. Boundary properties used for the simulations.